%% file: ANA-HDBS-2023-17-PAPER.tex
\documentclass[PAPER, atlasdraft=false, LANGEDIT=true, cernpreprint, UKenglish, texmf, orcidlogo]{atlasdoc}
 
\usepackage{atlaspackage}
\usepackage{atlasbiblatex}
 
\usepackage{atlasphysics}

\graphicspath{{logos/}{figures/}}
 
\usepackage{ANA-HDBS-2023-17-PAPER-defs}
 
\LEcontact{Bob Kowalewski bob.kowalewski@cern.ch}

\AtlasTitle{Combination of searches for resonant Higgs boson pair
production using $pp$ collisions at $\sqrt{s}=13$~$\text{Te\kern-0.1em V}$ with the ATLAS detector}

\AtlasAbstract{
A combination of searches for a new resonance decaying into a Higgs boson pair is presented,
using up to 139~fb$^{-1}$ of $pp$ collision data at $\sqrt{s}=13$~$\text{Te\kern-0.1em V}$
recorded with the ATLAS detector at the LHC.
The combination includes searches performed in three decay channels: $b\bar{b}b\bar{b}$, $b\bar{b}\tau^+\tau^-$
and $b\bar{b}\gamma\gamma$.
No excess above the expected Standard Model background is  observed and upper limits are set at the 95\% confidence
level on the production cross section of Higgs boson pairs
originating from the decay of a
narrow scalar resonance with mass in the range $251$~$\text{Ge\kern-0.1em V}$--$5$~$\text{Te\kern-0.1em V}$.
The observed (expected) limits are in the range 0.96--600~fb (1.2--390~fb).
The limits are interpreted in the Type-I Two-Higgs-Doublet Model
and the Minimal Supersymmetric Standard Model, and constrain
parameter space not previously excluded by other searches.}

\AtlasRefCode{HDBS-2023-17}
 
\PreprintIdNumber{CERN-EP-2023-271}

\AtlasJournal{Phys.\ Rev.\  Lett.}
\AtlasJournalRef{Phys. Rev. Lett. 132 (2024) 231801}
\AtlasDOI{10.1103/PhysRevLett.132.231801}

\hypersetup{pdftitle={ATLAS document},pdfauthor={The ATLAS Collaboration}}
 
\begin{document}

\maketitle

The discovery of a Higgs boson ($h$) with a mass of about \SI{125}{\GeV} at the Large Hadron Collider (LHC) in
2012~\cite{HIGG-2012-27,CMS-HIG-12-028}
has initiated a major effort
to understand the possible connections of this particle with new physics phenomena.
In particular, the possibility that
heavy particles beyond the Standard Model (SM) may couple to the
Higgs boson has attracted much attention.
This possibility is particularly compelling
because scalar
particles like the Higgs boson are dimension-one operators. Such operators allow
couplings between new particles and the Higgs boson that
are unsuppressed by higher energy scales~\cite{Patt:2006fw}.
These couplings could give rise to sizable effects evading indirect constraints from other searches.
Related signatures in which a heavy particle decays into a Higgs boson pair ($hh$)
are commonplace in almost all extensions of the Higgs sector, including models
with additional weak isospin singlets, doublets or
triplets~\cite{Abouabid:2021yvw,Branco:2011iw,Maniatis:2009re,Hartling:2014zca},
and in more exotic models, e.g. models with extra dimensions~\cite{Zhang:2015mnh}.
This type of resonant $hh$ production typically has a cross section that is several
orders of magnitude larger than the SM non-resonant $hh$ production~\cite{deFlorian:2016spz}, depending on the model.

This Letter presents the combination of searches for resonant Higgs boson pair production
in three $hh$ decay channels: four $b$-jets (\bbbb),
two $b$-jets and two $\tau$-leptons (\bbtt), and two $b$-jets and two photons (\bbyy).
These decay modes provide better sensitivity to Higgs boson pair production than other channels because they either have large Higgs boson branching ratios or
are easy to discriminate from background processes.
The searches under consideration use  the LHC Run-2 dataset corresponding to an
integrated luminosity of up
to 139\,fb$^{-1}$ of proton--proton ($pp$) collision data recorded with the ATLAS detector at a centre-of-mass
energy of $\sqrt{s}=13$~\TeV.
The ATLAS experiment at the LHC is a multipurpose particle detector with a forward--backward symmetric
cylindrical geometry and nearly $4\pi$ coverage in solid
angle~\cite{PERF-2007-01,ATLAS-TDR-2010-19,PIX-2018-001}, and uses an extensive software suite~\cite{ATL-SOFT-PUB-2021-001}
for both simulation and data analysis.
 
Previous combinations of searches for resonant Higgs boson pair production were performed using only part of the LHC Run-2 dataset corresponding to up to 36.1\,fb$^{-1}$ of $pp$ collision data.  The ATLAS Collaboration combined the decay modes \bbbb, \bbWW, \bbtt, \WWWW, \bbyy and \WWyy~\cite{HDBS-2018-58}, and the CMS Collaboration combined the decay modes \bbbb, \bbWW, \bbZZ, \bbtt\ and \bbyy~\cite{CMS-HIG-17-030}.  More recently, new searches were performed using the LHC Run-2 dataset in the \bbbb, \bbtt  and \bbyy decay channels by the ATLAS Collaboration~\cite{HDBS-2018-41,HDBS-2018-40,HDBS-2018-34}
and in the \bbbb,  \bbWW, \bbtt, \WWWW, \WWtt and \tttt decay channels
by the CMS Collaboration~\cite{CMS-B2G-21-003,CMS-B2G-20-007,CMS-HIG-21-002}.
These searches focus on a particle with mass greater than $250$~\GeV\ produced via gluon--gluon fusion
whose natural width is small compared with the experimental
mass resolution.  Gluon--gluon fusion is chosen because it is
typically a dominant production mechanism for this signature in most
extensions of the  Higgs sector.



Monte Carlo simulated signal samples for a new, narrow scalar resonance $X$ that is produced via gluon--gluon fusion
and decays into $hh$ ($X \rightarrow hh$),
are generated at leading
order accuracy in the strong coupling constant using the
\texttt{MADGRAPH5\_aMC@NLO} v 2.6.1 package~\cite{Alwall:2014hca} with the
\texttt{NNPDF2.3lo} parton distribution function set~\cite{Ball:2012cx}.
The event generator is interfaced with the \texttt{Herwig} v7.1.3
program~\cite{Bahr:2008pv,Bellm:2015jjp} to model the parton shower,
hadronisation and underlying event.
Different samples are generated in which the $X$ particle mass, \mX, is varied in the range $251\,\GeV$--$5\,\TeV$.
Its natural width is set to $10\,\MeV$, which is negligible compared with the experimental mass resolution, and the mass of the Higgs boson, $m_h$, is set to 125~\GeV.  The interference with non-resonant $hh$ production is neglected.
The interference is negligible for the combined cross-section upper limits that are presented in this Letter, due to the narrowness of the assumed $X$ resonance width. For the interpretation, neglecting the interference is an approximation.
Descriptions of the data and simulation samples
used to model backgrounds from SM processes can be found in Refs.~\cite{HDBS-2018-41,HDBS-2018-40,HDBS-2018-34}.


Each of the three 
decay channels included in this combination provides the highest sensitivity in the search for resonant $hh$ production for a portion of the \mX range considered.
The \bbbb final state has the highest branching ratio (33.9\%) and is the most sensitive channel for high resonance masses ($\mX \gtrsim 800$~\GeV), but the need to impose high \pt $b$-jet trigger thresholds, the ambiguity in the assignment of $b$-jets to Higgs boson  candidates, and the large multijet background reduce its sensitivity at lower \mX. The \bbtt  decay has a moderate branching ratio (7.3\%) and moderate background contamination and provides the best sensitivity in the intermediate resonance mass range ($350~\GeV \lesssim \mX \lesssim 800$~\GeV). The \bbyy final state has the lowest branching ratio (0.3\%), but benefits from the high trigger and reconstruction   efficiencies for photons. Combined with an excellent diphoton mass resolution, this channel has the best sensitivity at low resonance masses ($\mX \lesssim 350$~\GeV).
All three channels require the presence of $b$-quark initiated jets (\bjets), which are identified using the DL1r algorithm~\cite{FTAG-2018-01,FTAG-2019-07} at the 77\% efficiency working point. The DL1r algorithm is a deep-neural-network classifier that uses track and vertex information to classify the jet with respect to whether it contains $b$- or $c$-hadrons.  At the 77\% $b$-jet identification efficiency operating point,  light-jet ($c$-jet) rejection factors of 170 (5) are achieved in a sample of simulated SM \ttbar events.

The \bbbb analysis~\cite{HDBS-2018-41} is performed in two complementary channels: the resolved channel in which Higgs boson candidates are formed from four small-radius ($R=0.4$) \antikt jets~\cite{Cacciari:2008gp,Cacciari:2011ma}, and the boosted channel in which the high-\pt Higgs boson candidates ($\pt > 250$~\GeV) are reconstructed as two separate large-radius ($R=1.0$) \antikt jets.  The resolved channel relies on a combination of small-radius $b$-jet triggers~\cite{TRIG-2016-01, TRIG-2018-08} and uses \SI{126}{\ifb} of data, while the boosted channel uses large-radius jet triggers~\cite{TRIG-2016-01, TRIG-2019-02} without any $b$-tagging requirement applied and uses \SI{139}{\ifb} of data.
In the resolved channel, events are required to have at least four $R=0.4$ \antikt \bjets~\cite{PERF-2014-07,PERF-2015-09}, and a boosted decision tree (BDT) is used to pair the $b$-jets to form the two Higgs boson candidates.  In the boosted channel events are placed in the $2b$, $3b$ and $4b$ categories according to the number of $b$-tagged track-jets associated with each of the two $R=1.0$ \antikt jets that form the Higgs boson candidates.
These track-jets are built from inner-detector tracks clustered using the \antikt algorithm with a variable radius and are matched to the large-radius
jets using the ghost association method~\cite{Cacciari:2007fd},  as described in Ref.~\cite{HDBS-2018-41}.  The invariant masses of the two Higgs boson candidates are used to classify events into signal, validation, and control regions. The final observable in both channels is provided by the invariant mass of the two-Higgs-boson system, which is reconstructed with a resolution of $\sim 5$--$6$\%, thanks to the application in the analysis of a kinematic rescaling of the $h\to b\bar{b}$ system to the $h$-boson mass that improves the reconstructed $hh$ mass resolution.  The search is performed in the range $251\,\GeV< \mX <1.5\,\TeV$ ($900\,\GeV<\mX<5\,\TeV$) in the resolved (boosted) channel. The two channels are made orthogonal by explicitly vetoing resolved events that appear in the boosted selection, and are statistically combined in the overlapping mass region to enhance the sensitivity of the search.
 
The \bbtt~analysis~\cite{HDBS-2018-40} defines different event categories depending on
decay modes of $\tau$-leptons. The $\tau$-leptons decay into final states containing hadrons (\tauhad)
or leptons (\taulep)~\cite{ATLAS-CONF-2017-029,ATL-PHYS-PUB-2019-033}, and the analysis considers Higgs boson decays into \tauhadhad and \taulephad. The \tauhadhad channel relies on single-\tauhad and di-\tauhad triggers~\cite{TRIG-2016-01, TRIG-2019-02},  while the \taulephad channel uses single-lepton and lepton-plus-\tauhad triggers~\cite{TRIG-2016-01, TRIG-2019-02, TRIG-2018-01, TRIG-2018-05}. The search exploits the LHC Run-2 dataset of \SI{139}{\ifb}.  Events in the \tauhadlep channel are categorised in two orthogonal regions depending on whether they satisfy a single-lepton trigger or a lepton-plus-\tauhad trigger.  In the \tauhadhad category events are required to have exactly two reconstructed \tauhad candidates with opposite charge and no reconstructed electrons or muons, while in the \tauhadlep category events are required to have exactly one electron or muon and one \tauhad with opposite charge.  In both categories exactly two  $R=0.4$ \antikt \bjets are required.  The analysis uses the output score of a mass-parameterised neural network~\cite{Baldi:2016fzo} as the final observable in each signal region category.  The neural network can discriminate the mass of a possible signal excess with a resolution of 5--10\%, depending on \mX.
All \tauhadhad and \taulephad signal regions are statistically combined in the search for resonances in the range $251\,\GeV<\mX<1.6\,\TeV$.
 
The \bbyy analysis~\cite{HDBS-2018-34} relies on single-photon and diphoton triggers~\cite{TRIG-2016-01, TRIG-2018-05} and uses the LHC Run-2 dataset of 139~$\ifb$.  Events are required to have exactly two photons~\cite{EGAM-2018-01}, two $R=0.4$ \antikt \bjets, and zero electrons or muons.
Two BDTs are trained to separate the signal from backgrounds: one is trained against the $t\bar{t}\gamma\gamma$ background and another is trained against the single Higgs boson background. The signal region selection requires the mass of the diphoton system to be compatible with the SM Higgs boson mass and the mass of the $b\bar{b}\gamma\gamma$ system to be compatible with the expectation for the corresponding signal events. In addition, different requirements are applied on a combination of the two BDT scores with a tuned coefficient across the mass hypotheses.
For this combination, the \bbyy search follows the analysis strategy detailed in Ref.~\cite{HDBS-2018-34} but extends the $251\,\GeV<\mX<1.0\,\TeV$ mass range used there to $251\,\GeV<\mX<1.3\,\TeV$.
The final discriminating variable in this channel is the diphoton invariant mass, which is reconstructed with a resolution of about 1\%. The resolution of the mass of the $hh$ system in this channel is $\sim 2$--$3$\%, for the mass range considered
in this search, thanks to the excellent di-photon mass reconstruction and the use of a rescaling of the reconstructed $hh$ mass.


The results of the combination presented in this Letter are obtained from a likelihood function $L(\sigma,\vec{\theta})$, where $\sigma$ represents the parameter of interest of the model, namely the resonant $hh$ production cross section,  and $\vec{\theta}$ is a set of nuisance parameters, including the systematic uncertainty contributions and background parameters that are constrained by control regions in data.  The global likelihood function $L(\sigma,\vec{\theta})$ is obtained as the product of the likelihoods of the three searches included in the combination, which are themselves products of likelihoods computed from the final observables in the single analysis categories.
In this combination, the \bbbb, \bbtt and \bbyy analysis signal regions are either orthogonal due  to the different number and type of physics objects required in their final states, or have a negligible number of overlapping events.
They are therefore treated as statistically independent.
The profile-likelihood-ratio test statistic~\cite{Cowan:2010js} is used to obtain upper limits on the cross section of resonant $hh$ production with the CL$_\text{s}$ method~\cite{Read:2002hq}.  For signal masses up to 3~\TeV, the limits are computed using asymptotic formulae~\cite{Cowan:2010js}.  At higher masses, the asymptotic approximation is inaccurate due to the limited number of background candidates and the limits are computed by sampling pseudo-experiments.
 
Systematic uncertainties related to the data-taking conditions, such as those associated with the integrated luminosity and the modelling of the effect of multiple interactions in the same and neighbouring bunch crossings (pileup), are considered correlated across the searches. Uncertainties related to reconstructed quantities used by multiple searches and theoretical uncertainties on simulated signal and background processes are treated as correlated where appropriate.  Part of the $b$-tagging uncertainties and the signal parton shower modelling uncertainties are treated as uncorrelated due to different definitions and implementations in the individual analyses.  Different uncertainties dominate the combined limit in different \mX ranges.  These correspond to the dominant uncertainties of the channel most sensitive to resonant masses in that range. They are: photon scale and resolution, signal and background modelling from \bbyy for low masses~\cite{HDBS-2018-34}; $Z$ plus heavy-flavour jets modelling and single-Higgs boson plus heavy-flavour jets modelling from \bbtt for intermediate masses~\cite{HDBS-2018-40}; and multijet modelling, jet mass resolution and flavour-tagging from \bbbb for high masses~\cite{HDBS-2018-41}.  The total impact of the systematic uncertainties on the combined limit is between 3\% and 10\%, depending on \mX, and the impact of ignoring systematic uncertainty correlations between the three channels is of the order of 1\%.


The results of the combination are compatible with the predictions from
the SM backgrounds across the tested \mX range from 251\,\GeV to 5\,\TeV.  Figure~\ref{fig:pvalues} shows the local $p$-value
as a function of \mX for a narrow
resonance that decays into a pair of SM Higgs bosons.
The largest deviation is observed at 1.1\,\TeV and corresponds to a local significance of $3.3\,\sigma$, which is driven mainly by the \bbtt channel. The global significance of this excess is estimated to be $2.1\,\sigma$
using the method outlined in Ref.~\cite{Gross:2010qma}.
Figure~\ref{fig:limits} shows the upper limits at the 95\% confidence level (CL) on the resonant $hh$ production cross section as a function of \mX,
assuming that $h$ is the SM Higgs boson.
The observed (expected) upper limits are in the range
0.96--600~fb (1.2--390~fb), depending on \mX.
The \bbyy search is the most sensitive at low mass, the \bbtt search is
the most sensitive in the 350--800~\GeV\ range, and the \bbbb search dominates
for high resonant masses, demonstrating the complementarity
of these three searches.  These results represent an improvement in the upper limits by a factor
between two and five, depending on \mX, relative to the previous ATLAS combination~\cite{HDBS-2018-58}.
The improvement comes not only due to the use of a larger dataset but also due to improvements
in object identification and analysis techniques in background estimation.

\begin{figure}[htb]
\centering
\subfloat[]{
\includegraphics[width=0.5\textwidth]{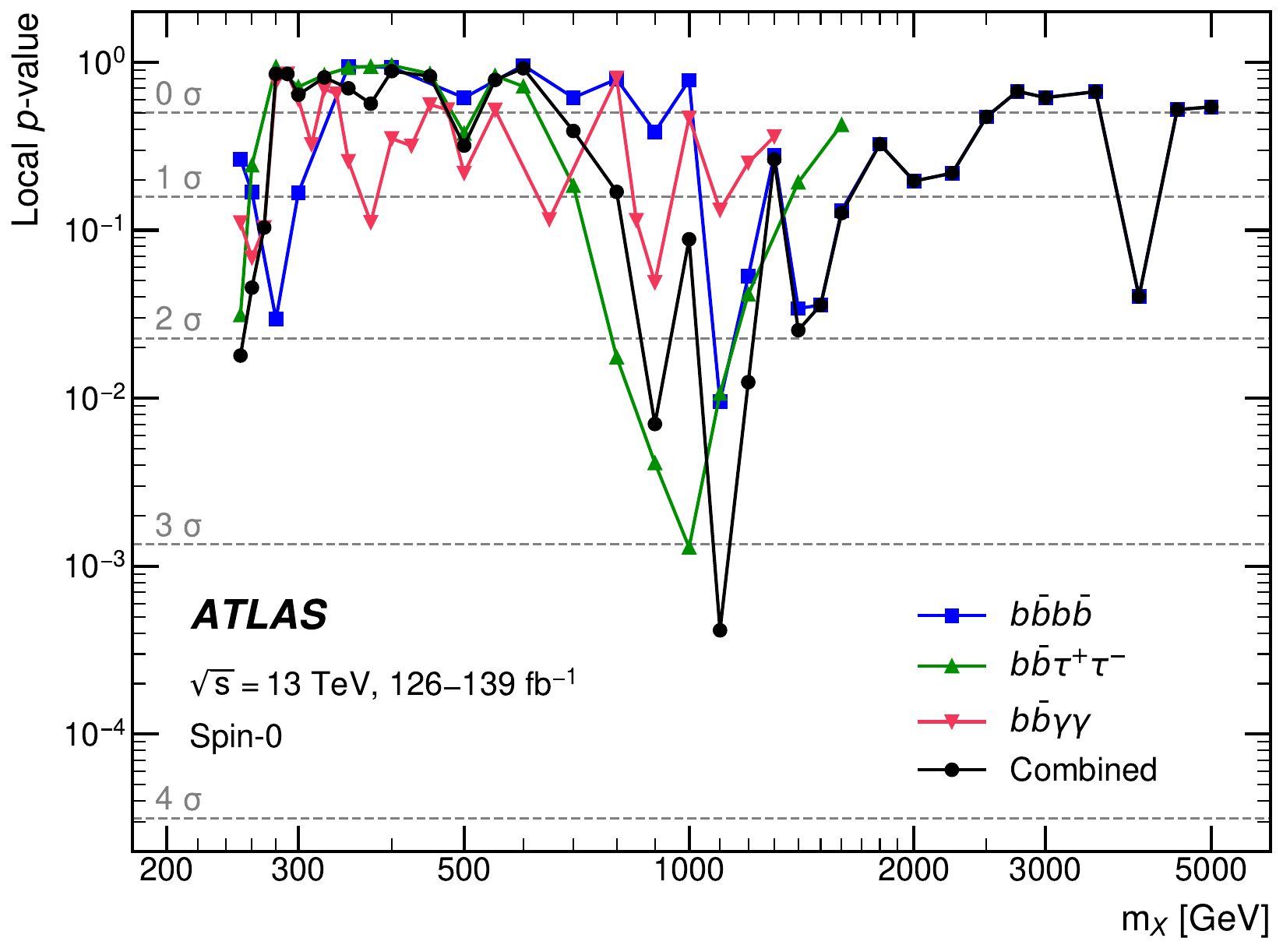}\label{fig:pvalues}}
\subfloat[]{
\includegraphics[width=0.5\textwidth]{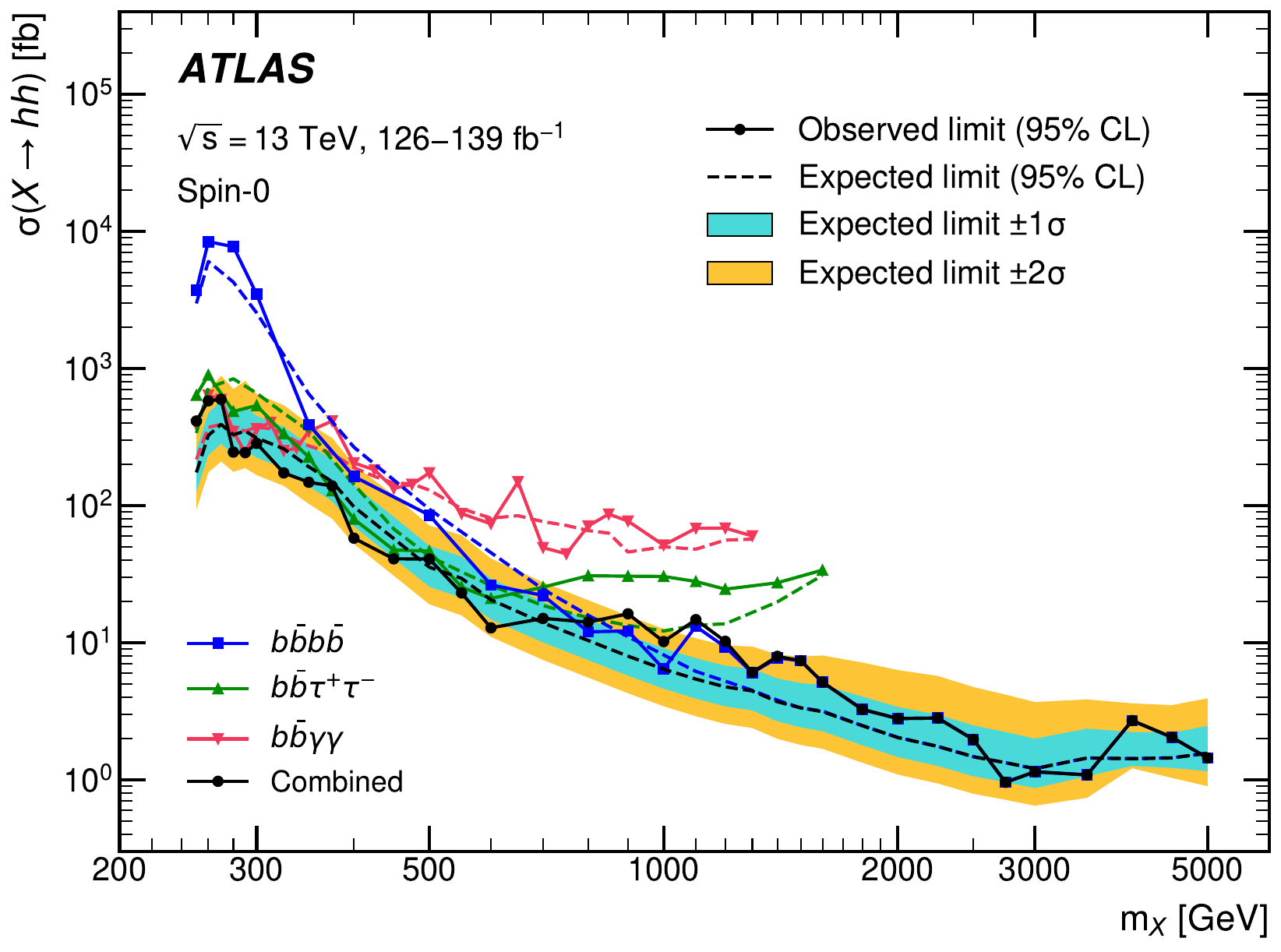}\label{fig:limits}}\\
\caption{(a) Local $p$-value and (b) observed and expected upper limits at the 95\% CL on the resonant Higgs boson pair production cross section as a function of the resonance mass \mX. The symbol $h$ denotes a SM Higgs boson with a mass 125\,\GeV.
}
\end{figure}


The upper limits are interpreted in the context of the Type-I Two-Higgs-Doublet
Model (2HDM) \cite{Lee:1973iz,Branco:2011iw} and the Minimal Supersymmetric SM
(MSSM) \cite{Fayet:1976et,Fayet:1977yc,Farrar:1978xj,Fayet:1979sa,Dimopoulos:1981zb}.
For these interpretations, the branching ratios of the Higgs boson are fixed to the predictions of the model under consideration.
 
The 2HDM is an extension of the SM with an additional Higgs doublet, leading
to five Higgs bosons after electroweak symmetry breaking,
three neutral and two charged.
The neutral Higgs bosons, assuming charge-parity (CP) conservation, can be CP-even ($h$
and $H$, with $m_H > m_h$) or CP-odd ($A$).
The resonant $hh$ production is assumed to come from $H\to hh$ for this interpretation.
The $h$ boson is assumed to be the Higgs boson observed at the LHC with $m_h=125$~\GeV.
The other Higgs bosons ($H$, $A$, $H^\pm$)
are assumed to be mass degenerate, i.e., $m_H=m_A = m_{H\pm}$, and the Higgs potential
parameter $m_{12}^2$ is fixed to $m_A^2 \tan\beta /(1+\tan^2\beta)$, where
$\tan\beta$ is the ratio of the vacuum expectation values of the two
Higgs doublets.  No constraint is imposed on $\cos(\beta-\alpha)$, where
$\alpha$ is the mixing angle between the CP-even Higgs bosons.  It,
along with $\tan\beta$ and $m_H$, is taken as a free parameter.
Widths and branching ratios
are calculated using the \texttt{2HDMC} program~\cite{Eriksson:2009ws}. The $H$ boson gluon--gluon fusion production
cross section calculation  uses the
\texttt{SusHi} package~\cite{Harlander:2012pb,Harlander:2016hcx}, which includes
corrections up to next-to-next-to-leading order in $\alpha_s$~\cite{Harlander:2002wh, Anastasiou:2002yz, Ravindran:2003um},
massive quarks~\cite{Harlander:2005rq,Spira:1995rr} and electroweak corrections by light
fermions~\cite{Aglietti:2004nj, Bonciani:2010ms}.
The procedure for the theoretical calculations follows Ref.~\cite{deFlorian:2016spz}.
The upper limits
on the Higgs boson pair production cross section
are interpreted as constraints on two benchmark planes:
$m_H$--$\tan\beta$ for  given $\cos(\beta-\alpha)$ values, shown in
Figures~\ref{fig:2hdm:p1:1}--\ref{fig:2hdm:p1:2}, and
$\cos(\beta-\alpha)$--$m_H$ for fixed $\tan\beta$ values, shown in
Figures~\ref{fig:2hdm:p2:1}--\ref{fig:2hdm:p2:2}.
The interpretation is given in the Type-I 2HDM in which gluon--gluon
fusion is the dominant production mechanism throughout the
parameter space where the searches have sensitivity.
The $H$ boson has a finite  natural width in the 2HDM. This is
taken into account by considering in the combination only channels for which
the experimental mass resolution is large in comparison to the $H$ boson natural width.
In particular, the \bbyy (\bbbb and \bbtt)  channel upper limits are valid for $H$ boson natural widths $\Gamma_H/m_H$
up to 2\% (5\%).
The results are sensitive
to $\cos(\beta-\alpha)$ values that are not probed by the SM
Higgs boson coupling measurements~\cite{HIGG-2018-57}.   
For example, the point $\tan\beta=10$,
$\cos(\beta-\alpha)=-0.1$ is excluded at 95\% CL for $m_H$ values in the range 270--810~\GeV, a region allowed by Higgs boson coupling measurements.

\begin{figure}[tbh]
\centering
\subfloat[]{
\includegraphics[width=0.5\textwidth]{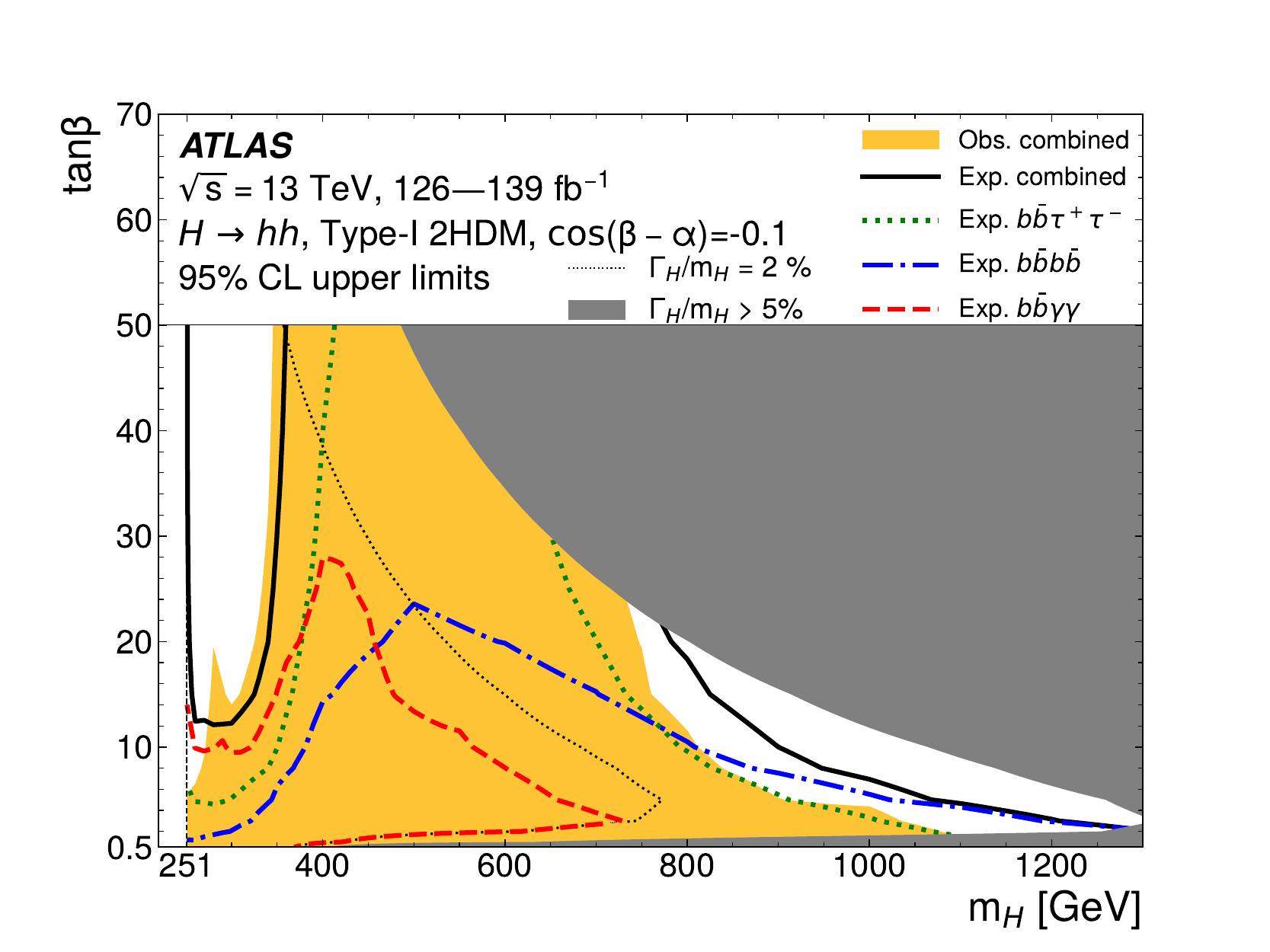}\label{fig:2hdm:p1:1} }
\subfloat[]{
\includegraphics[width=0.5\textwidth]{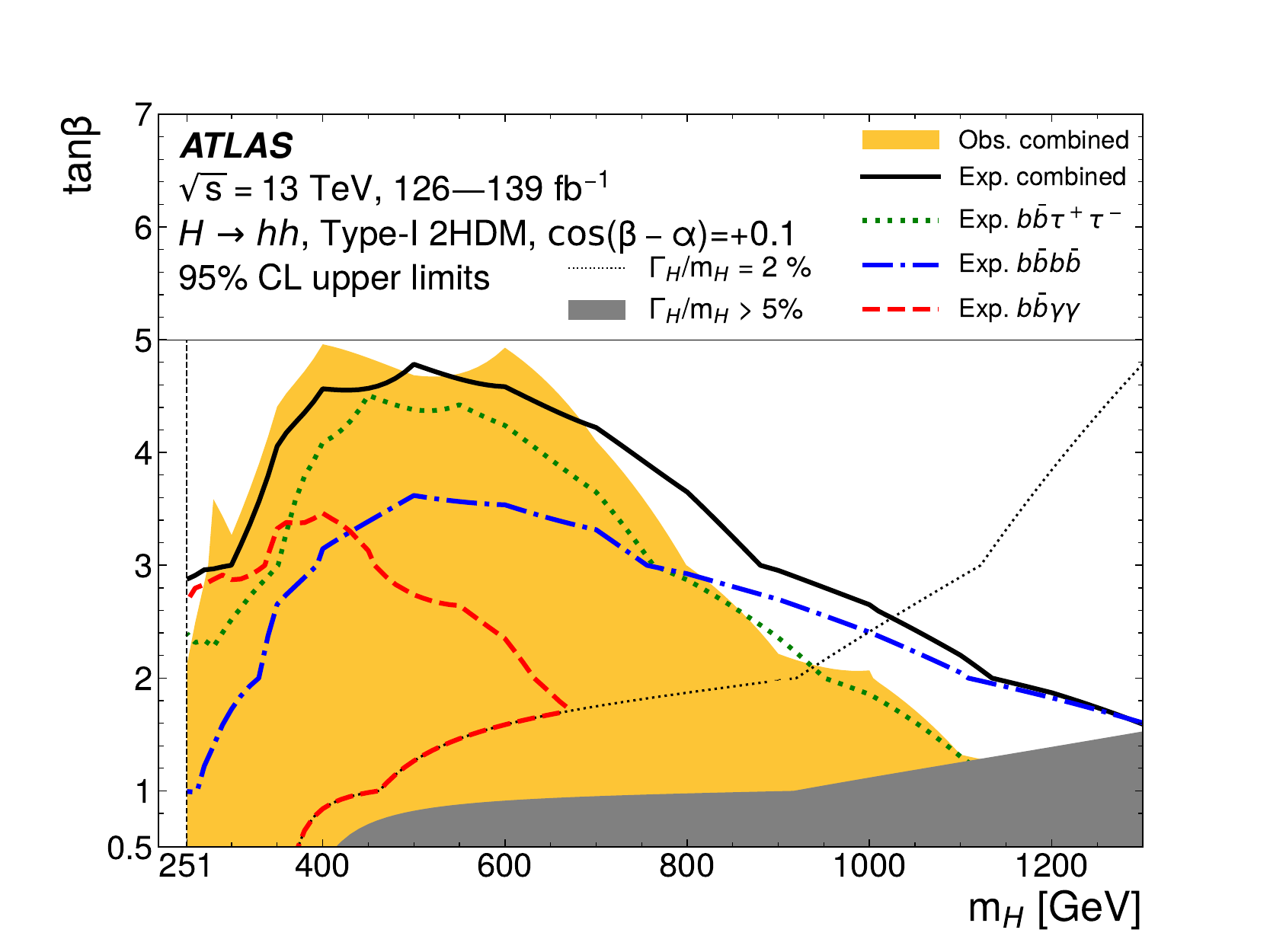}\label{fig:2hdm:p1:2}}\\
\subfloat[]{
\includegraphics[width=0.5\textwidth]{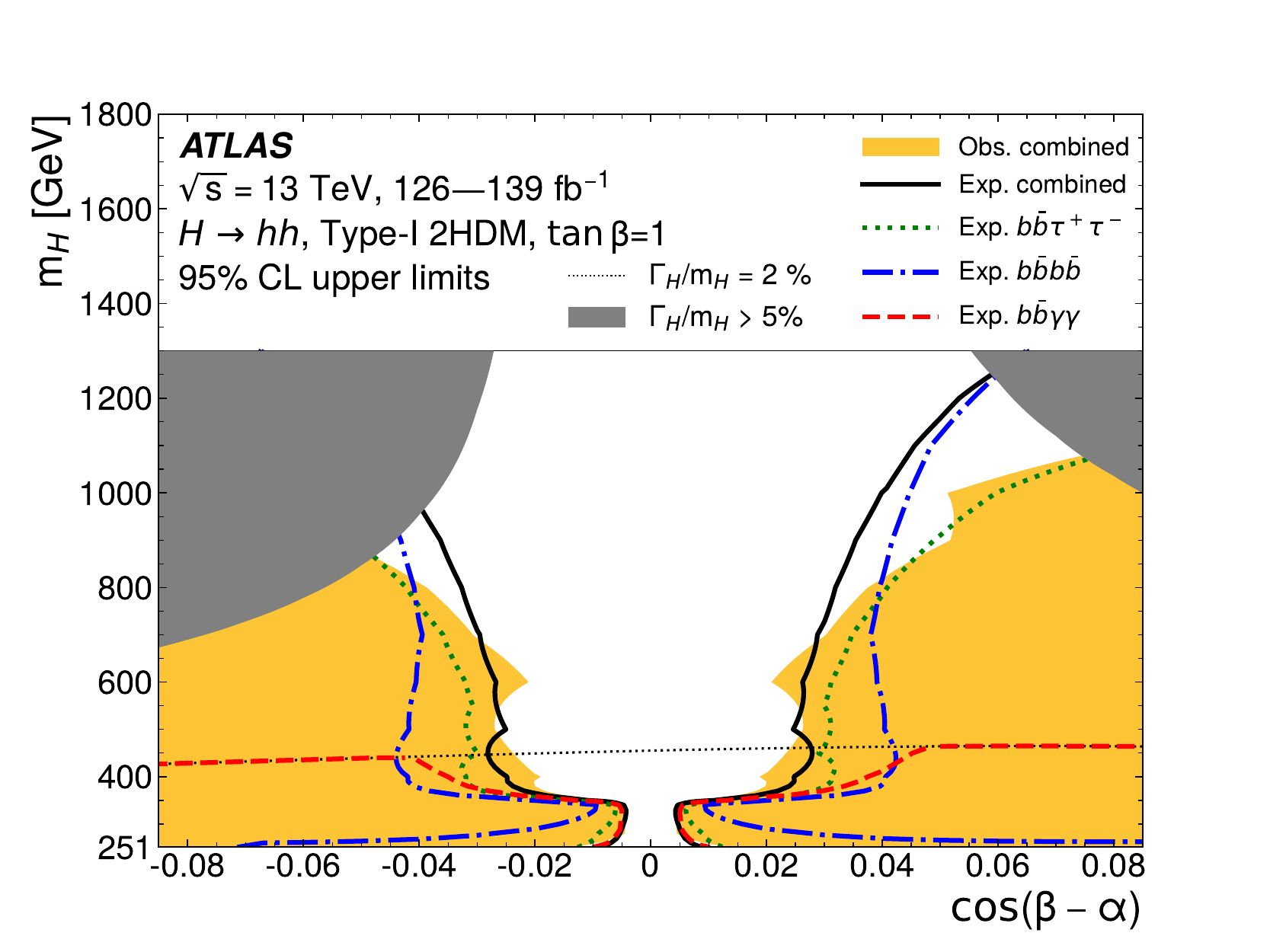}\label{fig:2hdm:p2:1}}
\subfloat[]{
\includegraphics[width=0.5\textwidth]{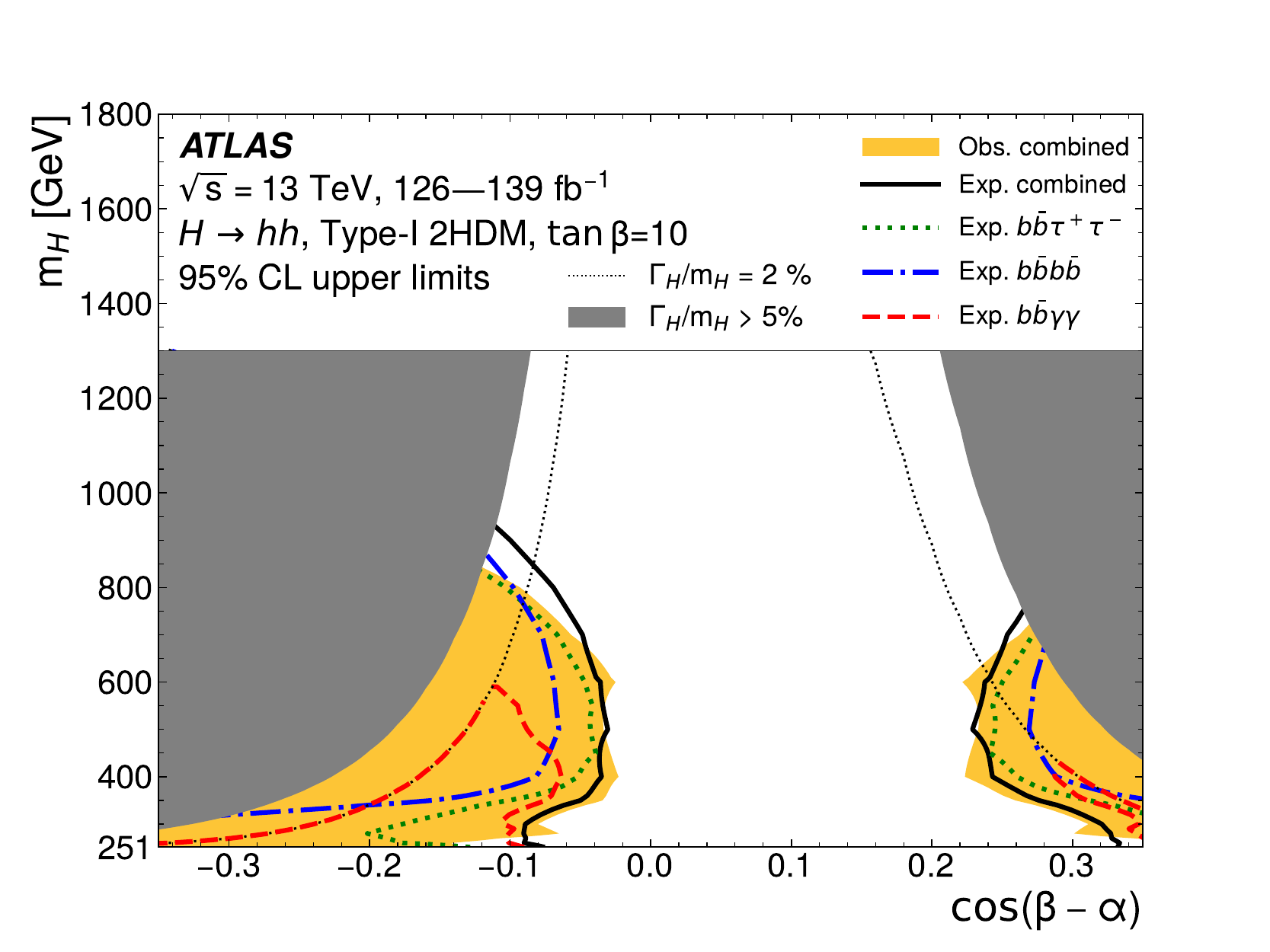}\label{fig:2hdm:p2:2}}
\caption{
Exclusion limits at the 95\% CL on the Type-I 2HDM parameter space.
Observed and expected limits for the combination of all channels and expected limits for each of the individual channels are presented.
The 2HDM parameters are shown in the $m_H$--$\tan\beta$ plane for (a) $\cos(\beta - \alpha)=-0.1$ and (b) $\cos(\beta - \alpha)=0.1$, and in the $\cos(\beta - \alpha)$--$m_H$ plane for (c) $\tan\beta=1$ and (d) $\tan\beta=10$.
The $\bbyy$ channel limits are valid for $H$ boson natural widths $\Gamma_H/m_H<2\%$, so limits from this channel are quoted for the parameter space that satisfies this requirement.
The limits in the $\bbbb$ and $\bbtt$ channels are only valid for $\Gamma_H/$m$_H<5\%$, so no limits are quoted in the dark shaded regions, where the width exceeds this value.
}
\end{figure}

The MSSM has a Type-II 2HDM Higgs sector structure~\cite{Djouadi:2005gj} and,
therefore, includes the 2HDM parameters discussed earlier.
The resonant $hh$ production is assumed to come from $H\to hh$ for this interpretation.
Supersymmetry constrains the number of free parameters in the Higgs
sector at lowest order to be two, taken here as $m_A$ and $\tan\beta$.
Radiative corrections have a large impact on the MSSM and the limits
are influenced by how the supersymmetry parameters are chosen, with
each choice defining a particular scenario. In this
Letter, the M$^{125}_{h,\text{EFT}}$ and M$^{125}_{h,\text{EFT}}(\tilde\chi)$ scenarios
are used~\cite{Bahl:2019ago}.
These scenarios have supersymmetry
mass parameters that are not related to the Higgs sector at a very high energy scale
such that the low $\tan\beta$ region has $m_h \approx 125$~\GeV.
The M$^{125}_{h,\text{EFT}}(\tilde\chi)$ scenario differs from M$^{125}_{h,\text{EFT}}$
in that it includes low-mass neutralinos and charginos.
The MSSM gluon--gluon fusion cross section is obtained using a \texttt{SusHi}
implementation that includes supersymmetric QCD
corrections~\cite{Degrassi:2010eu,Degrassi:2012vt}. The
Higgs boson masses and mixing are calculated with the \texttt{Feynhiggs}
program~\cite{Heinemeyer:1998yj, Heinemeyer:1998np, Degrassi:2002fi, Frank:2006yh, Hahn:2013ria, Bahl:2016brp, Bahl:2017aev,Bahl:2018qog}
as described in Ref.~\cite{Bahl:2018jom}.
The Higgs boson branching ratio calculation uses \texttt{FeynHiggs}, \texttt{HDECAY}
and \texttt{PROPHECY4f} as described in Refs.~\cite{LHCHiggsCrossSectionWorkingGroup:2011wcg,Dittmaier:2012vm}.
The upper limits on the Higgs boson pair production cross section
are interpreted in the MSSM as constraints in the $m_A$--$\tan\beta$ plane, as shown in Figure~\ref{fig:bsm:EFT}.
The combined analysis excludes parameter space in the region $2 \lesssim \tan\beta \lesssim 5$,
which is not excluded by the $H/A\to\tau\tau$, $A\to Zh$, $H\to ZZ$ or $H^{\pm}\to tb$
searches~\cite{HDBS-2018-46,HDBS-2020-19,HIGG-2018-09,HDBS-2018-51}.
 
\begin{figure}[tbh]
\centering
\subfloat[]{
\includegraphics[width=0.5\textwidth]{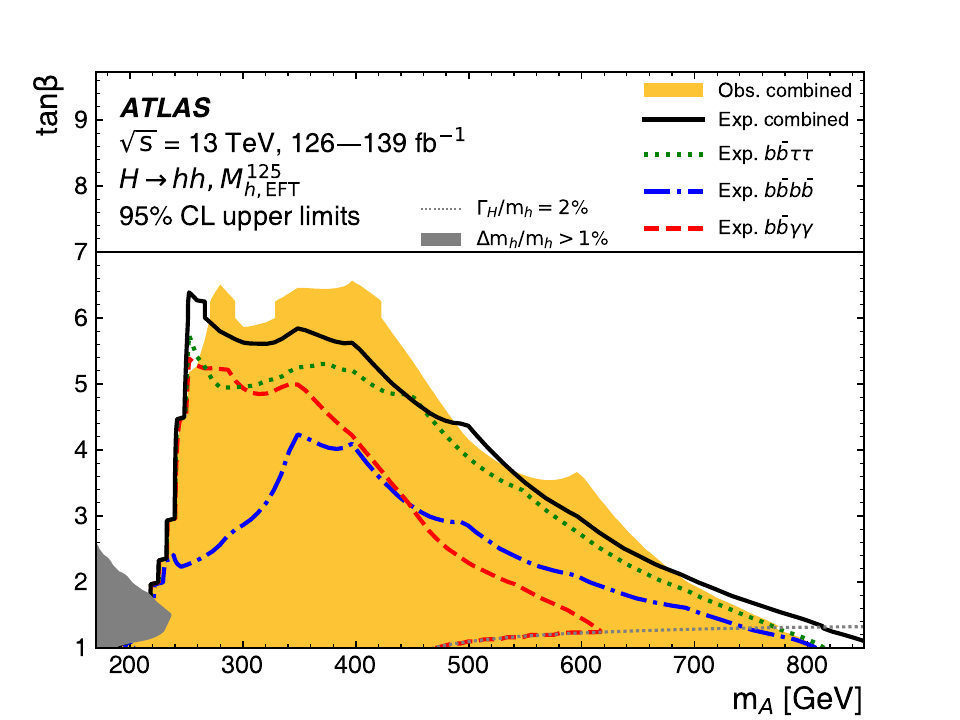}}
\subfloat[]{
\includegraphics[width=0.5\textwidth]{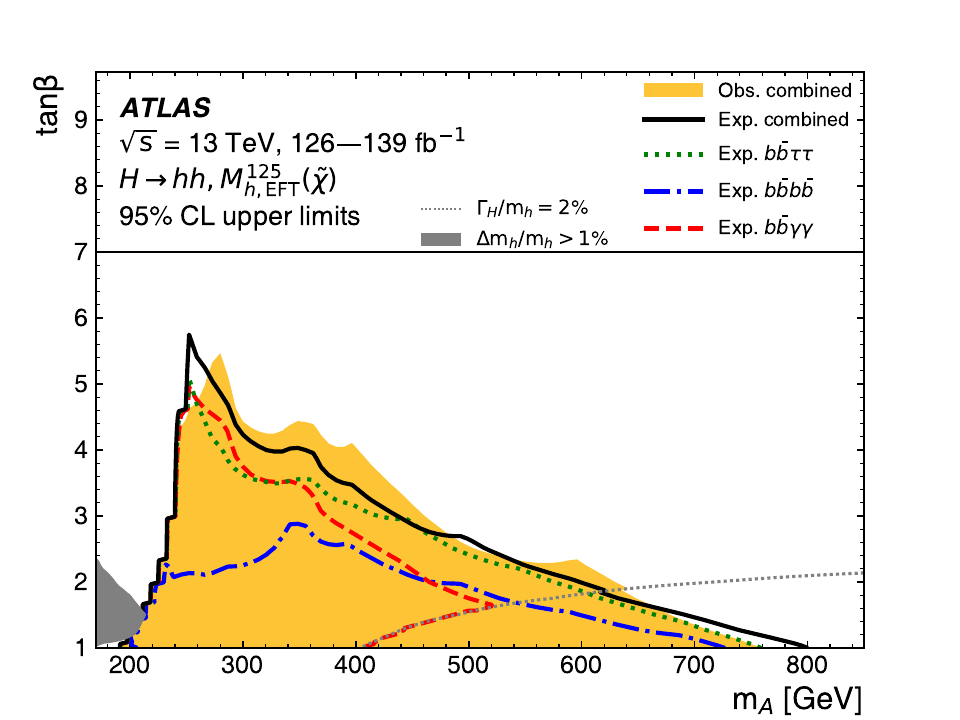}}
\caption{Observed and expected exclusion limits at the 95\% CL on the MSSM
parameter space for the (a) M$^{125}_{h,\text{EFT}}$ and
(b) M$^{125}_{h,\text{EFT}}(\tilde \chi)$
benchmark scenarios
for each of the individual channels and their combination.
The $\bbyy$ channel limits are valid for $H$ boson natural widths $\Gamma_H/m_H<2\%$, so limits from this channel are quoted for the parameter space that satisfies this requirement.
The limits do not apply in the dark shaded regions, where the mass of
the lightest CP-even Higgs boson, $h$, is not compatible with 125\,\GeV
within the experimental mass resolution.
}
\label{fig:bsm:EFT}
\end{figure}


In conclusion, this Letter reports a combined search for a narrow-width resonance decaying to $hh$ in the \bbbb, \bbtt and \bbyy final states using up to 139~fb$^{-1}$ of $pp$ collision data at $\sqrt{s}=13$~\TeV\ collected by the ATLAS experiment.  The data are found to be consistent with the SM prediction and upper limits at the 95\% CL are set on the resonant $X \rightarrow hh$ cross section assuming \mX to be in the range 251~\GeV--5~\TeV. The observed (expected) upper limits are in the range 0.96--600~fb (1.2--390~fb) and they constitute an improvement of a factor of 2--5, depending on \mX, with respect to the previous ATLAS combined result~\cite{HDBS-2018-58}.  The results are also interpreted in the context of the Type-I 2HDM and MSSM, excluding parameter space that was hitherto
allowed by the most sensitive searches for these models.


\section*{Acknowledgements}


We thank CERN for the very successful operation of the LHC and its injectors, as well as the support staff at
CERN and at our institutions worldwide without whom ATLAS could not be operated efficiently.
 
The crucial computing support from all WLCG partners is acknowledged gratefully, in particular from CERN, the ATLAS Tier-1 facilities at TRIUMF/SFU (Canada), NDGF (Denmark, Norway, Sweden), CC-IN2P3 (France), KIT/GridKA (Germany), INFN-CNAF (Italy), NL-T1 (Netherlands), PIC (Spain), RAL (UK) and BNL (USA), the Tier-2 facilities worldwide and large non-WLCG resource providers. Major contributors of computing resources are listed in Ref.~\cite{ATL-SOFT-PUB-2023-001}.
 
We gratefully acknowledge the support of ANPCyT, Argentina; YerPhI, Armenia; ARC, Australia; BMWFW and FWF, Austria; ANAS, Azerbaijan; CNPq and FAPESP, Brazil; NSERC, NRC and CFI, Canada; CERN; ANID, Chile; CAS, MOST and NSFC, China; Minciencias, Colombia; MEYS CR, Czech Republic; DNRF and DNSRC, Denmark; IN2P3-CNRS and CEA-DRF/IRFU, France; SRNSFG, Georgia; BMBF, HGF and MPG, Germany; GSRI, Greece; RGC and Hong Kong SAR, China; ISF and Benoziyo Center, Israel; INFN, Italy; MEXT and JSPS, Japan; CNRST, Morocco; NWO, Netherlands; RCN, Norway; MEiN, Poland; FCT, Portugal; MNE/IFA, Romania; MESTD, Serbia; MSSR, Slovakia; ARRS and MIZ\v{S}, Slovenia; DSI/NRF, South Africa; MICINN, Spain; SRC and Wallenberg Foundation, Sweden; SERI, SNSF and Cantons of Bern and Geneva, Switzerland; MOST, Taipei; TENMAK, T\"urkiye; STFC, United Kingdom; DOE and NSF, United States of America.
 
Individual groups and members have received support from BCKDF, CANARIE, CRC and DRAC, Canada; CERN-CZ, PRIMUS 21/SCI/017 and UNCE SCI/013, Czech Republic; COST, ERC, ERDF, Horizon 2020, ICSC-NextGenerationEU and Marie Sk{\l}odowska-Curie Actions, European Union; Investissements d'Avenir Labex, Investissements d'Avenir Idex and ANR, France; DFG and AvH Foundation, Germany; Herakleitos, Thales and Aristeia programmes co-financed by EU-ESF and the Greek NSRF, Greece; BSF-NSF and MINERVA, Israel; Norwegian Financial Mechanism 2014-2021, Norway; NCN and NAWA, Poland; La Caixa Banking Foundation, CERCA Programme Generalitat de Catalunya and PROMETEO and GenT Programmes Generalitat Valenciana, Spain; G\"{o}ran Gustafssons Stiftelse, Sweden; The Royal Society and Leverhulme Trust, United Kingdom.
 
In addition, individual members wish to acknowledge support from CERN: European Organization for Nuclear Research (CERN PJAS); Chile: Agencia Nacional de Investigaci\'on y Desarrollo (FONDECYT 1190886, FONDECYT 1210400, FONDECYT 1230812, FONDECYT 1230987); China: National Natural Science Foundation of China (NSFC - 12175119, NSFC 12275265, NSFC-12075060); Czech Republic: PRIMUS Research Programme (PRIMUS/21/SCI/017); European Union: European Research Council (ERC - 948254, ERC 101089007), Horizon 2020 Framework Programme (MUCCA - CHIST-ERA-19-XAI-00), European Union, Future Artificial Intelligence Research (FAIR-NextGenerationEU PE00000013), Italian Center for High Performance Computing, Big Data and Quantum Computing (ICSC, NextGenerationEU); France: Agence Nationale de la Recherche (ANR-20-CE31-0013, ANR-21-CE31-0013, ANR-21-CE31-0022), Investissements d'Avenir Labex (ANR-11-LABX-0012); Germany: Baden-Württemberg Stiftung (BW Stiftung-Postdoc Eliteprogramme), Deutsche Forschungsgemeinschaft (DFG - 469666862, DFG - CR 312/5-2); Italy: Istituto Nazionale di Fisica Nucleare (ICSC, NextGenerationEU); Japan: Japan Society for the Promotion of Science (JSPS KAKENHI JP21H05085, JSPS KAKENHI JP22H01227, JSPS KAKENHI JP22H04944, JSPS KAKENHI JP22KK0227); Netherlands: Netherlands Organisation for Scientific Research (NWO Veni 2020 - VI.Veni.202.179); Norway: Research Council of Norway (RCN-314472); Poland: Polish National Agency for Academic Exchange (PPN/PPO/2020/1/00002/U/00001), Polish National Science Centre (NCN 2021/42/E/ST2/00350, NCN OPUS nr 2022/47/B/ST2/03059, NCN UMO-2019/34/E/ST2/00393, UMO-2020/37/B/ST2/01043, UMO-2021/40/C/ST2/00187, UMO-2022/47/O/ST2/00148); Slovenia: Slovenian Research Agency (ARIS grant J1-3010); Spain: BBVA Foundation (LEO22-1-603), Generalitat Valenciana (Artemisa, FEDER, IDIFEDER/2018/048), La Caixa Banking Foundation (LCF/BQ/PI20/11760025), Ministry of Science and Innovation (MCIN \& NextGenEU PCI2022-135018-2, MICIN \& FEDER PID2021-125273NB, RYC2019-028510-I, RYC2020-030254-I, RYC2021-031273-I, RYC2022-038164-I), PROMETEO and GenT Programmes Generalitat Valenciana (CIDEGENT/2019/023, CIDEGENT/2019/027); Sweden: Swedish Research Council (VR 2018-00482, VR 2022-03845, VR 2022-04683, VR grant 2021-03651), Knut and Alice Wallenberg Foundation (KAW 2017.0100, KAW 2018.0157, KAW 2018.0458, KAW 2019.0447); Switzerland: Swiss National Science Foundation (SNSF - PCEFP2\_194658); United Kingdom: Leverhulme Trust (Leverhulme Trust RPG-2020-004); United States of America: U.S. Department of Energy (ECA DE-AC02-76SF00515), Neubauer Family Foundation.


\printbibliography
 
\clearpage
\input{atlas_authlist}

\end{document}

%% file: atlas_authlist.tex
 
\begin{flushleft}
\hypersetup{urlcolor=black}
{\Large The ATLAS Collaboration}

\bigskip

\AtlasOrcid[0000-0002-6665-4934]{G.~Aad}$^\textrm{\scriptsize 102}$,
\AtlasOrcid[0000-0001-7616-1554]{E.~Aakvaag}$^\textrm{\scriptsize 16}$,
\AtlasOrcid[0000-0002-5888-2734]{B.~Abbott}$^\textrm{\scriptsize 120}$,
\AtlasOrcid[0000-0002-1002-1652]{K.~Abeling}$^\textrm{\scriptsize 55}$,
\AtlasOrcid[0000-0001-5763-2760]{N.J.~Abicht}$^\textrm{\scriptsize 49}$,
\AtlasOrcid[0000-0002-8496-9294]{S.H.~Abidi}$^\textrm{\scriptsize 29}$,
\AtlasOrcid[0000-0002-9987-2292]{A.~Aboulhorma}$^\textrm{\scriptsize 35e}$,
\AtlasOrcid[0000-0001-5329-6640]{H.~Abramowicz}$^\textrm{\scriptsize 151}$,
\AtlasOrcid[0000-0002-1599-2896]{H.~Abreu}$^\textrm{\scriptsize 150}$,
\AtlasOrcid[0000-0003-0403-3697]{Y.~Abulaiti}$^\textrm{\scriptsize 117}$
% ...
\bigskip
\\

% ...
$^{16}$Department for Physics and Technology, University of Bergen, Bergen; Norway.\\
% ...
$^{29}$Physics Department, Brookhaven National Laboratory, Upton NY; United States of America.\\
% ...
$^{35}$$^{(a)}$Facult\'e des Sciences Ain Chock, R\'eseau Universitaire de Physique des Hautes Energies - Universit\'e Hassan II, Casablanca;$^{(b)}$Facult\'{e} des Sciences, Universit\'{e} Ibn-Tofail, K\'{e}nitra;$^{(c)}$Facult\'e des Sciences Semlalia, Universit\'e Cadi Ayyad, LPHEA-Marrakech;$^{(d)}$LPMR, Facult\'e des Sciences, Universit\'e Mohamed Premier, Oujda;$^{(e)}$Facult\'e des sciences, Universit\'e Mohammed V, Rabat;$^{(f)}$Institute of Applied Physics, Mohammed VI Polytechnic University, Ben Guerir; Morocco.\\
% ...
$^{49}$Fakult\"{a}t Physik , Technische Universit{\"a}t Dortmund, Dortmund; Germany.\\
% ...
$^{55}$II. Physikalisches Institut, Georg-August-Universit\"{a}t G\"ottingen, G\"ottingen; Germany.\\
% ...
$^{102}$CPPM, Aix-Marseille Universit\'e, CNRS/IN2P3, Marseille; France.\\
% ...
$^{117}$Department of Physics, New York University, New York NY; United States of America.\\
% ...
$^{120}$Homer L. Dodge Department of Physics and Astronomy, University of Oklahoma, Norman OK; United States of America.\\
% ...
$^{150}$Department of Physics, Technion, Israel Institute of Technology, Haifa; Israel.\\
$^{151}$Raymond and Beverly Sackler School of Physics and Astronomy, Tel Aviv University, Tel Aviv; Israel.\\
% ...
$^{a}$ Also Affiliated with an institute covered by a cooperation agreement with CERN.\\
$^{b}$ Also at An-Najah National University, Nablus; Palestine.\\
$^{c}$ Also at Borough of Manhattan Community College, City University of New York, New York NY; United States of America.\\
$^{d}$ Also at Center for High Energy Physics, Peking University; China.\\
$^{e}$ Also at Center for Interdisciplinary Research and Innovation (CIRI-AUTH), Thessaloniki; Greece.\\
$^{f}$ Also at Centro Studi e Ricerche Enrico Fermi; Italy.\\
% ...

\end{flushleft}
